\documentclass{aa}  

\usepackage{graphicx}
\usepackage{txfonts}
\def\simlt{\lower.5ex\hbox{$\; \buildrel < \over \sim \;$}}
\def\simgt{\lower.5ex\hbox{$\; \buildrel > \over \sim \;$}}
\def\simpropto{\lower.2ex\hbox{$\; \buildrel \propto \over \sim \;$}}

\begin{document}
   \title{A giant bar induced by a merger event at z=0.4?}

\author{S. Peirani\inst{1},
F. Hammer\inst{2}, H. Flores\inst{2}, Y. Yang\inst{2} \and E. Athanassoula\inst{3}}

    \institute{Institut d'Astrophysique de Paris, 98 bis Bd Arago, 75014
 Paris, France
% - Unit\'e mixte de recherche 7095 CNRS - Universit\'e Pierre et Marie Curie\\
  - UMR 7095 CNRS - Universit\'e Pierre et Marie Curie\\
             \email{peirani@iap.fr}
       	\and
     GEPI, Observatoire de Paris, CNRS,
     University Paris Diderot, 5 place Jules Janssen,
     92190 Meudon, France\\% \email{}
         \and
     LAM, UMR6110,
     CNRS/Universit\'e de Provence, Technop\^ole de Marseille-Etoile, 
     38 rue Fr\'ed\'eric Joliot Curie, 13388 Marseille C\'edex 20, France \\ 
             }

   \date{Received ..., ...; accepted ..., ...}

% \abstract{}{}{}{}{} 
% 5 {} token are mandatory
 
  \abstract
   {
    Disk galaxies are the most frequent objects among the galaxy
    population in the local universe. However, the formation of their
    disks and substructures --  in particular their bars -- is still a
    matter of debate. 
   }
   {
    We present a physical model of the formation of
    J033239.72-275154.7, a galaxy observed at $z=0.41$ and characterized 
    by a big young bar of size $6$ kpc. 
    The study of this system is particularly interesting for understanding
    the connection between mergers and bars as well as the properties and fate of
    this system as it relates to disk galaxy formation.
    %It is probable that this
    %object is the progenitor of a present day spiral and therefore its
    %study is fundamental to understand and to constraint the galaxy disk
    %formation process.
   }
   {
    We compare the morphological and kinematic properties of
    J033239.72-275154.7, the latter obtained by the GIRAFFE
    spectrograph, to those derived from 
    the merger of two spiral galaxies described by idealized N-body simulations
    including a star formation prescription.
   }
   {
    We found that the general morphological shape and most of the dynamical
    properties of the object can be well reproduced by a model in which
    the satellite is initially put in a retrograde orbit and the
    mass ratio of the system is 1:3.
    In such a scenario, a bar forms in the host galaxy after the
    first passage of the satellite
    where an important fraction of available gas is consumed in an
    induced burst. In its later evolution, however, we find that
    J033239.72-275154.7, whose major progenitor was an Sab galaxy,
     will probably become a S0 galaxy.
    %and not a disk
    This is mainly due to the violent relaxation and the angular momentum loss
    experienced by the host
    galaxy during the merger process, which is caused by the adopted
    orbital parameters.
    This result suggests that the
    building of the Hubble sequence is significantly influenced by  the last
    major collision.  In the present case, the merger leads to a
    severe damage of the disk of the progenitor, leading to an evolution
    towards a more bulge dominated galaxy. 
 %The merger event enhances the star formation but not
  %  destroy entierly the disk: i n its later evolution, our model suggest also that
   % J033239.72-275154.7 would probably
   % become a galaxy with a thin disk.
   %  This supports the idea that disk galaxies may be more frequent 
   % to survive to major mergers.   
   % However, some remaining discrepancies between our numerical
   % modelling and the observations tend to suggest that even distant galaxies
   % may present morphological and dynamical properties too constraint.
   }
   {%The combination of dynamical properties derived from GIRAFFE, the
   %star formation history and the morphology impose very strong
   %constraints and rule out most of candidates based on morphological
   %criteria only. Moreover, even at $z=0.4$ observations are too more
   %constraints to be modelled by simple N body simulations.
  }


  \maketitle

%
%________________________________________________________________

\section{Introduction}

The formation of disk galaxies remains an outstanding puzzle in 
contemporary astrophysics (see Mayer et al. 2008 for a review).
According to hierarchical models of structure formation, mergers and
interaction of galaxies are an essential ingredient of galaxy formation and
evolution. Earlier works and numerical simulations show that the
remnants of mergings of purely stellar progenitors are more
likely to be elliptical galaxies 
(Toomre 1977; Barnes 1988; Barnes \& Hernquist 1992; Hernquist 1992; Lima-Neto 
\& Combes 1995; Balcells \& Gonz{\'a}lez 1998; Naab et al. 1999 etc.)
 and recent studies extended this result to gas-rich
  progenitors (Springel \& Hernquist 2005; Robertson
et al. 2006; Hopkins et al. 2008). Whether this is compatible, or
incompatible, with the fraction of disk galaxies present in the local
universe depends on the typical number
of expected mergers by galaxy (see for instance Kazantzidis et al. 2007).
Such predictions seems to be incompatible with observations
which suggest that disk galaxies represent the majority (70\%) of the galaxy
population observed in the local universe
(see Hammer et al. 2005; Nakamura et al. 2004 and
references therein).
To help resolve such issues,
Hammer et al. (2005) suggested that
disks can be rebuilt during the encounters of gas rich spirals.
Indeed such proposition was guided by the remarkable coincidence
of the redshift increase, up to z=1, of the merger
rate, of the fraction of actively star forming galaxies (including
luminous IR galaxies, LIRGs), of the fraction of galaxies with
peculiar galaxies (including those with compact morphologies)
etc... This has been followed by simulations of gas rich mergers
(Springel \& Hernquist 2005; Robertson
et al. 2006; Governato et al. 2007; Hopkins et al. 2008),
% Robertson et
%al. (2006); Governato et al. (2006),
evidencing that under certain
conditions, a disk may be re-built after a merger. Indeed such a
conclusion had been already reached by Barnes (2002), but these
simulations are less convincing since
they do not include a prescription for star formation.
% More recently
Lotz et al. (2006) have also analysed a large suite of simulated
equal mass rich mergers and find that most merger remnants appear
disc-like and dusty. 
% Hammer et al. (2005) proposed an
%alternative scenario in which 
%a large fraction of local disks could have been rebuilt after a major
%merger event.  Such a suggestion is supported by the coincidence
%between the number evolution with redshift of mergers, star forming
%galaxies (including LIRGs) and of galaxies with peculiar morphologies.
%In fact, this conclusion was already reached from the
%simulations of Barnes (2002) and confirmed recently  with 
%new high resolution simulations of the merger of isolated
%gas rich galaxies with star formation recipes (Springel \& Hernquist 2005; Robertson
%et al. 2006; Governato et al. 2007; Hopkins et al. 2008).
Moreover, this scenario is also consistent with results of   
 other simulations and semi-analytical
models which claim that, without merger processes, most of galaxies and their host dark
matter halos cannot acquire the required angular momentum to form disks
(see Peirani et al. 2004; Puech et al. 2007 and references therein).  
The formation of bars is also a fundamental issue in the evolution of
disk galaxies, particularly since it has now been shown that only
about one third of them were in place at $z$ = 0.8 (Sheth et al. 2008).

Our aim here is to reproduce with
numerical modelling the general morphology (presence
of a bar, of substructures, etc...),
the dynamical properties of the gas component (velocity fields),
the photometric properties of stars (colors, star formation rate,
etc...) of J033239.72-275154.7, a galaxy located at $z=0.41$ for which
we have high quality imaging and kinematics. This work will also provide
useful input to disk formation models, since it gives information on
potential progenitors of the present-day galaxy disks, as well as
constraints on their formation 
(initial orbital configuration, mass ratio of the system, etc...)
and possible fate. 
This work is an offspring of a VLT large program entitled IMAGES
(``Intermediate-MAss Galaxy Evolution Sequence'', Yang et al. 2008) which
gathers high quality kinematics for a representative sample of $\sim
100$ massive galaxies at 
$z=0.4-0.75$ and with $M_J(AB)\leq-20.3$. Using the GIRAFFE
spectrograph at the VLT, the kinematic properties of 65 of these
galaxies, J033239.72-275154.7 for instance, have been derived.  
%We propose to study here one of them, called J033239.72-275154.7.
This galaxy lies at $z=0.41$, is
classified as a merger from analysis lead by Neichel et al. (2008)
and presents a big, young bar. 
This bar has a size of 6 kpc,  i.e. is quite big 
since 72\% of barred galaxies from a  sample of
2000 galaxies from the SDSS have a size lower than this
 (Barazza et al. 2008).
% given the fact that
%in the systematic study of bars in a sample of
%2000 galaxies from the SDSS (Barazza et al. 2008) find that
%72\% of barred galaxies have a size lower than 6 kpc.
This bar has also an extremely blue color, consistent
with a starburst, i.e. with ages well below few
100s of Myr. This contrasts with many bars in the local universe which are
known to include relatively old and
red stars (Sheth et al. 2003).
This paper is organized as follows: in Section 2 we present the general
morphological and kinetic properties of J033239.72-275154.7 and 
in Section 3 a short description of
the numerical modelling and of the results obtained. We conclude  
in Section 4.

\section{General properties of J033239.72-275154.7}
\label{sect_GP}

J033239.72-275154.7 has been presented first by Yang et al. (2008).
Table \ref{table0} summarizes its overall properties, including 
photometry, morphological parameters and kinematics measurements
(provided by the IMAGES data basis and can be retrieved from
Yang et al. 2008, Neichel et al. 2008 and Puech et al. 2008).
This object has a stellar mass of $2.0\times 10^{10} M_\odot$, a
K-band magnitude of $M_K=-20.94$
and shows a peculiar morphology and
kinematics. Its center is dominated
by an elongated structure, most likely a giant thin bar of semi-major
axis 0.85 arcsec or
about 6 kpc. 
Neichel et al. (2008) found that its color is roughly
 b-z=0.8, i.e. typical of starburst (see their figures 8 and
 12).
 % since its color value, derived 
 %in Neichel et al. 2008, is b-z=0.8. 
 %The typical color is b-z=0.8
 %(see Figure 8 of Neichel et al. 2008) and this value has
 %been derived from this paper (color map description and Figure 12).
This bar is embedded into a diffuse region, which is probably a disk,
as argued e.g. by features which look like tidal arms (one very blue
at the upper right and another on the left). It is, nevertheless, very
irregular, with several blue clumps. Roughly at its center, there is a
strong light condensation which we will sometimes refer to as the
core, and which is probably a small bulge
% (color from late type to
%Sbc)
Its color is roughly b-z=1.8, i.e. as found in galaxies of
  late type to Sbc (Neichel et al. 2008).
The bar is asymmetric, extending spatially much more towards the
upper-right side of the core.
On the other side of the core, the bar is redder and a part of it disappears
at UV wavelengths.
The discrepancy in color between the two sides is roughly
  0.3 magnitudes, as shown by the CDFS-GOODS observations 
  (b band observations, thus UV rest-frame at z=0.4). 
 On the bottom of the galaxy, there are two bright
adjacent knots, which dominate the rest-frame UV light. These knots have
the color of a pure starburst and could be hardly resolved. Due to the
irregularity of the disk and to the existence of the bright tidal
arms, it is difficult to measure accurately its 
position and inclination angle. From the shape of the outer isophotes
we find that the photometric major axis is roughly
in the upper-right lower-left direction and the inclination should lie
between $28^\circ$ and $35^\circ$.
The star formation rate (SFR) is moderate ($SFR_{IR}=6.6$,
$SFR_{2800}=4.2$) and
the dust does not affect much the overall color.
%the dust is not affecting  much the overall color.
Note that
SFRs have been estimated using the 2800A and 15 micron luminosities, respectively. Those have
been converted to SFR using the method described in Kennicutt (1998). Luminosities have been
estimated using interpolations between observed photometric points provided by HST/ACS,
EIS and Spitzer/MIPS, using the method described in Hammer et al. (2001 \& 2005). The ratio SFR
of the SFR is consistent with Av=0.22 (standard extinction curve) which is very low. 
Finally, the electron density is modest,
including at the very blue knots.

%%%%%%%%%%%%%%%%%%%%%%%%%%%%%%%%%%%%%%%%%%%%%%%%%
%                   FIGURE ONE
%%%%%%%%%%%%%%%%%%%%%%%%%%%%%%%%%%%%%%%%%%%%%%%%%
\begin{figure}
\rotatebox{0}{\includegraphics[width=3.6cm]{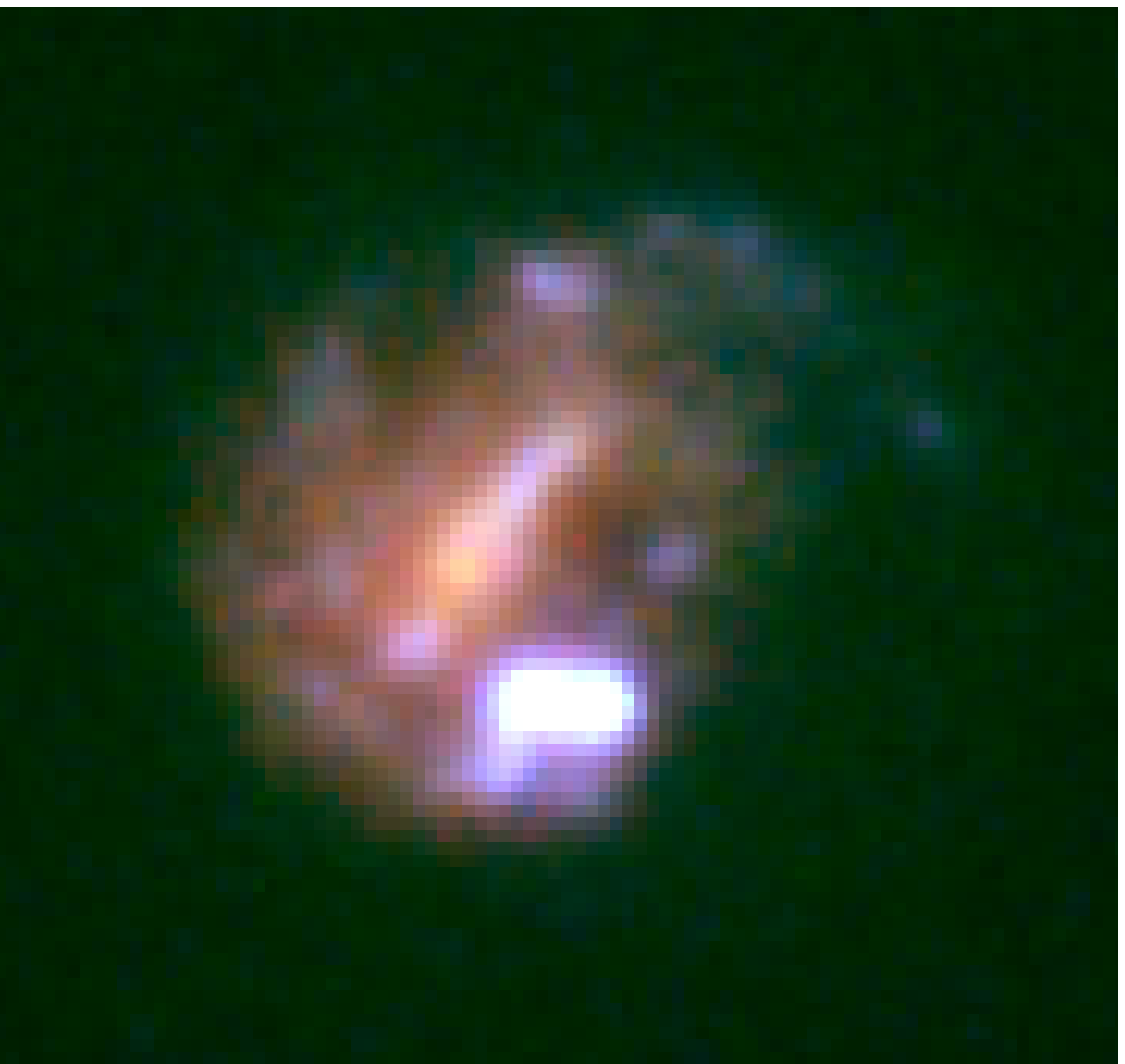}}
\rotatebox{0}{\includegraphics[width=5.1cm]{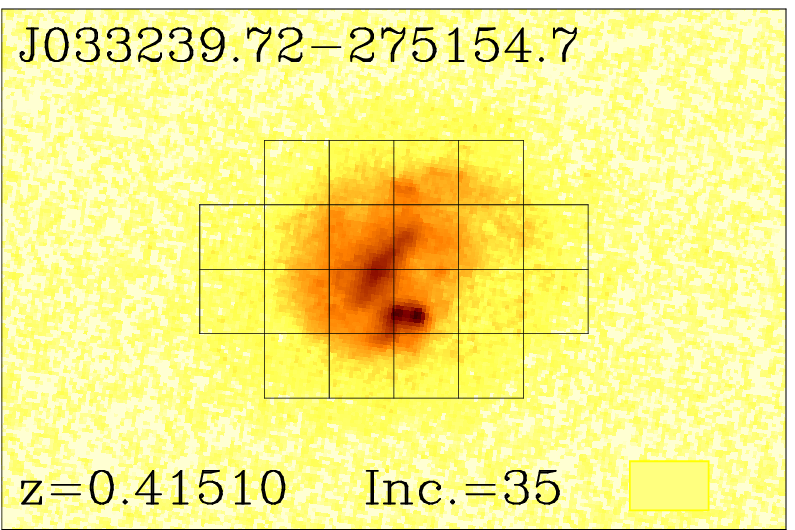}}\\[0.2cm]
\rotatebox{0}{\includegraphics[width=4.4cm]{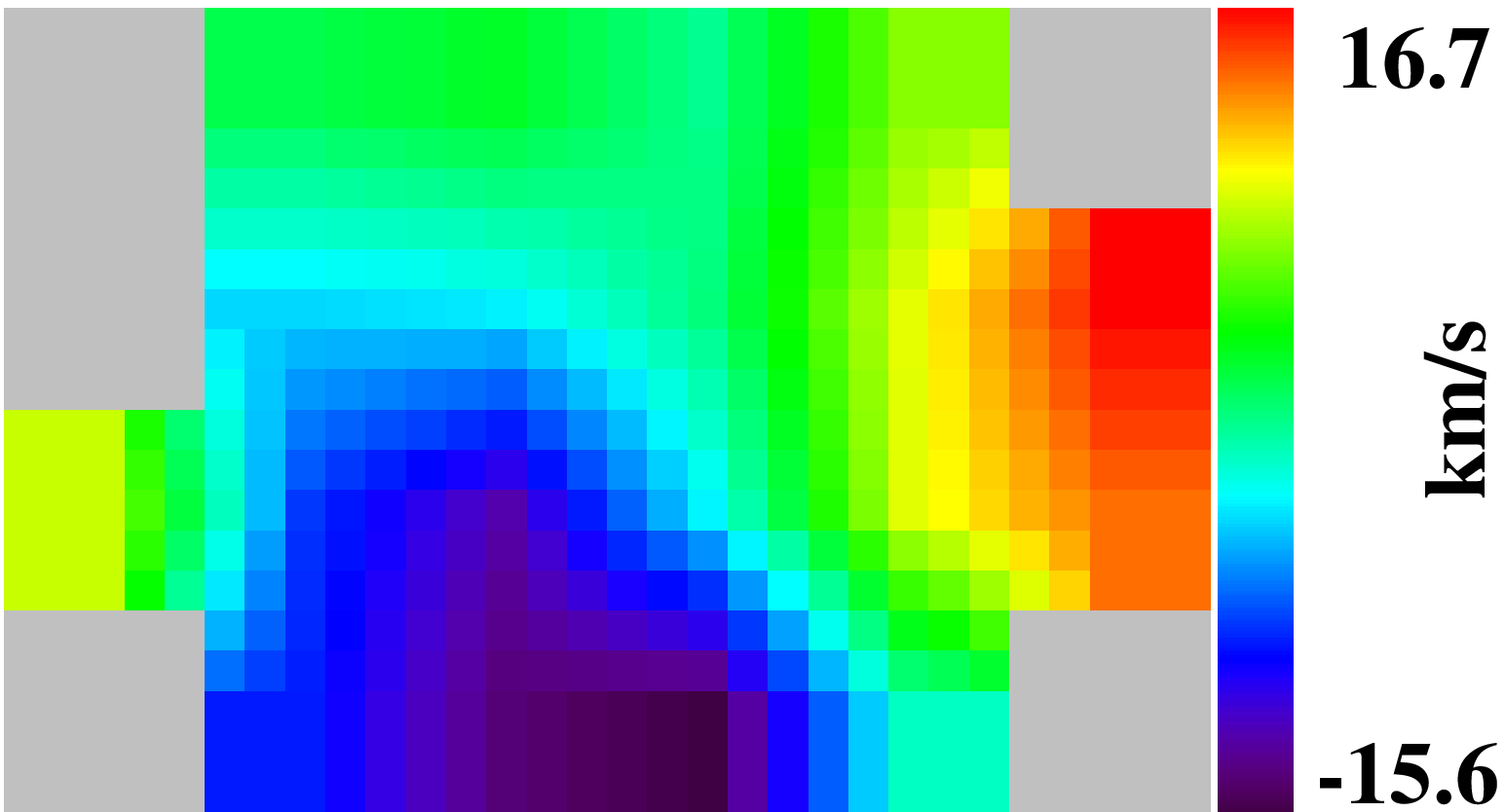}}
\rotatebox{0}{\includegraphics[width=4.4cm]{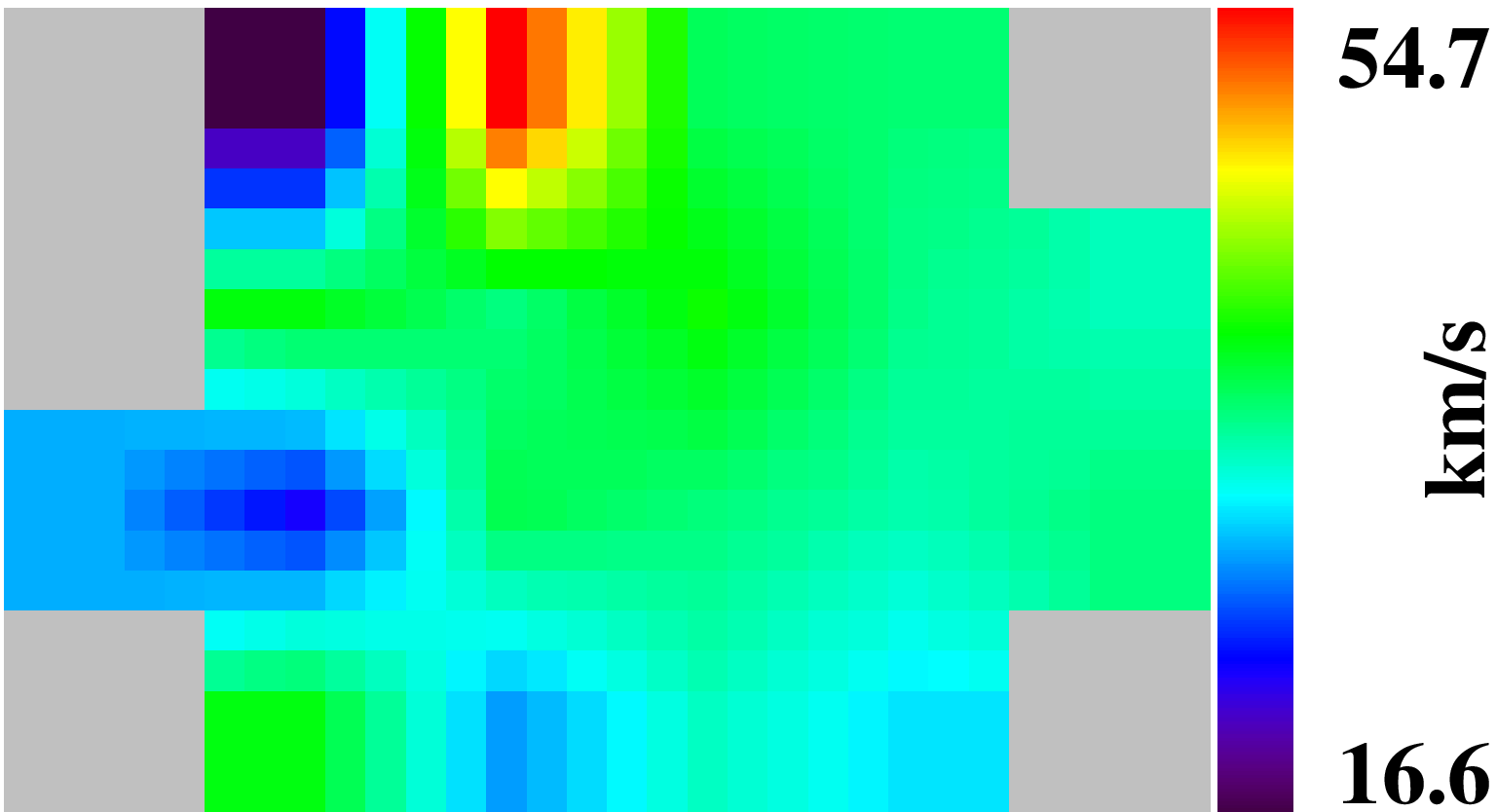}}
\caption{B-V-z color map of J033239.72-275154.7
from HST/ACS (upper left panel) and distribution of the
galaxy within 
  the GIRAFFE grid (upper right panel). The lower panels show the
 velocity field map (lower left panel) and the associated $\sigma$-map (lower right panel).
 }
 \label{vf_sig}
 \end{figure}
%%%%%%%%%%%%%%%%%%%%%%%%%%%%%%%%%%%%%%%%%%%%%%%%%%

\begin{table}
\caption{Properties of J033239.72-275154.7}             
% Neuf col alignes a droite (l)
{\scriptsize
\begin{tabular}{llllllll}\hline
\multicolumn{8}{l}{Multi wavelength photometry, stellar mass and SFR}\\\hline
 M$_{2800}$ & M$_{B}$ & M$_{V}$ &M$_{J}$  & M$_{K}$ & $M_{stellar}$ &  SFR$_{2800}$  &SFR$_{IR}$ \\
& & &  & &log(M$_{\odot}$) & M$_{\odot}$/yr & M$_{\odot}$/yr   \\
-19.56 &-20.10 &-20.62& -21.04& -20.94 &10.31 &4.2   &6.6  \\\hline
\multicolumn{8}{l}{Morphology and Kinematics}\\
\multicolumn{8}{l}{
(from GIRAFFE measurements, Puech et al. 2008; Yang et al. 2008)}\\\hline
V$_{flat}$ & r$_{half}$ &   $incl_{disk}$ & size of the bar  \\
km/s &  kpc & deg.  &  kpc   \\
30 & 3.5  &35.4  & $\sim 6$  \\\hline
\end{tabular}
}
\label{table0}
\end{table}

The velocity field (VF) is obviously complex (see Fig. \ref{vf_sig}).
The kinematical major axis is almost parallel to the bar and it is
offset by more than one GIRAFFE pixel from the bulge,
towards the prominent blue knots (in the bottom). 
The pixel scale is 0.52 arcsec and the full width half
  maximum spectral resolution
 is 23 km/s. Details on the GIRAFFE
instruments can be found in Flores et al. (2006) and Yang et al. (2008).
The small VF
amplitude is dominated by the two star-bursting clumps and by the blue
arm/giant tidal tail. Even so, the velocity
amplitude is relatively small, of the order of 30 km/s. The velocity
dispersion map is nearly featureless at a value of 30-40 km/s over most
of the galaxy. It has, nevertheless, a clear, localised sigma peak
(50 km/s), which is offset by 2 GIRAFFE pixels from the bulge and could coincide
(to within half a GIRAFFE pixel) with a small and relatively blue clump on the
upper-right located roughly where the blue tidal arm joins the disk. 
The signal to noise ratio in this pixel is quite high ($\sim 50$).

All the evidence argues that strong interaction and/or merging is at
work in this system. Indeed, the dynamical axis is off-centered and
passes through the region where the two bright knots lie.
Moreover, there are strong tidal arms, and 
the value of $V_{rot}$/$\sigma$ is small (about 1-2 depending on the
exact value of the inclination). Indeed, the velocity dispersion is
about 30-40 km/s over most of the galaxy. Given the VF values and the
measured inclination,  the rotational velocity is far 
 below the value expected from the Tully-Fisher (TF) relation, which is 
 $V_{rot}=125$ km/s. This value has been taken from Hammer et al. (2007)
who have carefully estimated the various samples used to derive this
relationship. 
 Given the spatial resolution, it is not easy to derive a
  rotation curve. Nevertheless, thorough modelling of the
velocity fields have been done by Puech et al. (2008), who investigate the evolution of
the Tully Fisher relation. 
Given its kinematical and morphological properties, it is thus unlikely
that J033239.72-275154.7 can be a rotating disk 
 hosting two giant HII regions because: (1) the two knots have a z band
luminosity corresponding to one third of the total z luminosity
(the relative photometry has been done using apertures, scaled to the
  full width half maximum of
the knot luminosity, using the IRAF/DAOPHOT/ package) 
 ;
 (2) 
the dynamical axis is strongly off-centered; (3) the value of $\sigma$
is high and that of $V_{rot}$ low.

Could merging of the two bright knots with
  J033239.72-275154.7 be compatible with the formation of a 
giant bar with a relatively blue color? $N$-body simulations have shown
that a small companion merging with a disk galaxy could, depending on
its density and its orbit, either destroy a pre-existing bar
(Athanassoula 1999; Berentzen et al. 2003), or trigger its formation
(Walker et al. 1996; Berentzen et al. 2004).

%%%%%%%%%%%%%%%%%%%%%%%%%%%%%%%%%%%%%%%%%%%%%%%%%
%                   FIGURE TWO
%%%%%%%%%%%%%%%%%%%%%%%%%%%%%%%%%%%%%%%%%%%%%%%%%
\begin{figure*}
\rotatebox{0}{\includegraphics[width=18cm]{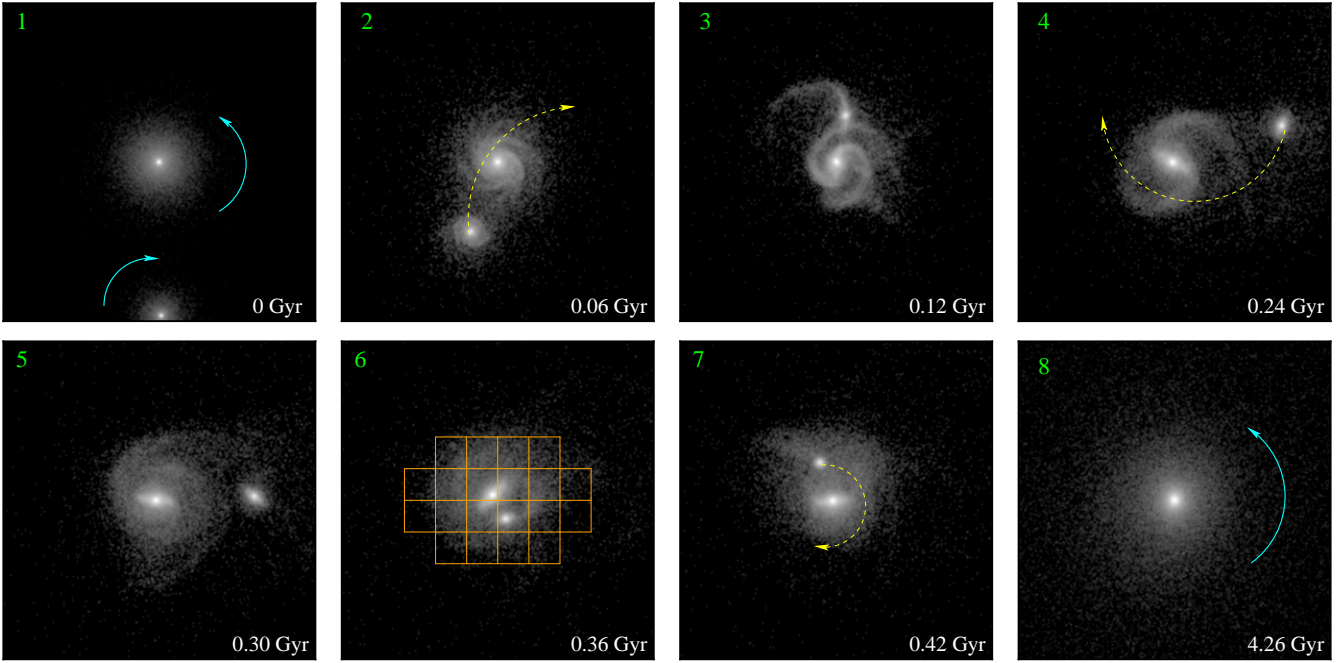}}
%\rotatebox{0}{\includegraphics[width=4.4cm]{galastar.ps}}
%\rotatebox{0}{\includegraphics[width=4.4cm]{galaxy_sim2.ps}}
%\rotatebox{0}{\includegraphics[width=4.4cm]{galaxy_sim3.ps}}
\caption{Time evolution of the projected star number density. The 
 light blue arrows indicate the specific rotation of each
 galaxy, while the yellow dashed lines show the motion of the satellite.
 In panel 6, we superposed
 the GIRAFFE grid. Each frame is 40 kpc $\times$ 40 kpc in size.
}
\label{simulated_galaxy}
 \end{figure*}
%%%%%%%%%%%%%%%%%%%%%%%%%%%%%%%%%%%%%%%%%%%%%%%%%

\section{Simulations}

We briefly describe the numerical methodology used to
model J033239.72-275154.7. We use idealized N-body simulations of
the merger of two spiral galaxies.
One of the most difficult part of this work results from the
huge number of free parameters in the initial conditions, for instance
related to the
mass ratio and the orbital configuration of the system or 
the star formation prescription. We have
performed more than 100 simulations and we present in this section our
fiducial modelling. First, as mentioned in section \ref{sect_GP}, the 
dynamical axis of the system is off-centered and 
the velocity field strongly indicates that the
satellite is falling down toward the main object. Then, by trying
different orbital configurations, we found that retrograde orbits
for the satellite are 
more plausible regarding the orientation of the tidal
arms of the main galaxy (in the left and upper right parts)
and the specific position of the accreted satellite. Second,
different mass ratios of the system,  different inclinations between the
two orbit planes and different pericentric distances 
have been tested at the same time in order to match the amplitudes of the
observed VF and $\sigma$ map. For instance, since the VF amplitudes are
small, the inclination  between the
two orbital planes should be small.
It is also worth mentioning that we restricted our study 
to parabolic orbits (Khochfar \& Burkert 2006) to reduce
the number of free parameters. 
Finally, different parameters related to the star formation
recipes (see below) have also been tested in order to obtain results consistent
with the observations.

\subsection{Initial conditions and numerical method}

Our galaxies are composed of a spherical dark matter halo (with a
Hernquist profile, Hernquist 1990), a disk, composed of stars and gas,
and a bulge. In both objects, the disk and
the bulge represent respectively 15\% and 5\% of 
the total mass.
The baryon fraction used here is slightly higher than the
cosmic baryon fraction ($\sim 16.7$\%) derived by Komatsu et al. 2008 in 
order to compare to and match the results from Barnes (2002). 
It's worth mentioning that the gas fraction in the disk is 25\% in the
host galaxy, whereas it is 50\% in the accreted satellite. 
 These gas fractions are consistent with the estimation of Liang et
 al. (2006)  Gavazzi et al. (2008) and Rodrigues et
   al. (2008). Note that the Gavazzi et al. (2008)
 measurements have been made in the local universe, while Liang et
 al. (2006) using the M-Z relationship, were able to 
estimate the evolution of the gas phase
in z=0.6 galaxies, and found that,
on average, galaxies at z=0.6 have two times more gas than at present
days.   
The host galaxy has a stellar mass of $1.3 \times 10^{10} M_{\odot}$ and
is more likely a Sab galaxy.
%The host galaxy has a total mass of $8.25 \times 10^{10} M_{\odot}$ and
%the mass ratio of the system is 1:3.
 Galaxies are created following
Springel et al. (2005).
 Dark matter haloes have a concentration parameter of
$C_{host}=14$ and $C_{sat}=15$  for the host and the
satellite, respectively, in good agreement with Dolag et al. (2004).
 Their spin
parameter, defined by $\lambda ={J\mid E \mid ^{1/2}}/{GM^{5/2}}$
where $J$ is the angular momentum, $E$ is the
total energy of the halo and $M$ is its mass,
 is $\lambda_{host}=\lambda_{sat}=0.05$ in order that
the rotation curves of the galaxies are closed to the baryonic
TF relation (McGaugh 2005). Indeed, their
maximum circular velocity is 140 km/s and 98 km/s, respectively.  
In our fiducial model, the satellite is
put on a retrograde parabolic orbit (Khochfar \& Burkert 2006) 
with a pericentric distance
$R_{p}=2.1$ kpc  and initial separation of $20$
kpc. The inclination between the two orbit planes is $15^\circ$ and
the galaxy spins are opposed. 
 Note that the initial separation between the two galaxies is quite
small in order to reduce the high computational cost of all
experiments. A higher distance would decrease 
the halos overlap at the beginning of the simulation, but this should not
change the main results and conclusions of this paper.

%\subsection{Numerical method}
The simulation is performed using GADGET2 
(Springel 2005) with added prescriptions for cooling, star
formation and feedback from Type Ia and II supernovae (SN).
Approximately $275,000$ particles are used for the experiment and the
masses ($M$), gravitational softening lengths ($\epsilon$) and
number of particles ($N$) of each component 
involved are summarized in table\ref{table1}.

\begin{table}
\begin{center}
\caption{Masses ($M$), gravitational softening lengths ($\epsilon$) and
number of particles ($N$) of each component used in the simulations.} 
\begin{tabular}{c|c|c|c|c}
\hline
 & DM & gas & star (disk) & star (bulge)\\
\hline
%Mass ($10^{5} M_\odot$) & 6.600 & 0.6445 & 3.609 & 1.719\\
$M_{host}$ ($10^{10} M_\odot$) & 6.60 & 0.31 & 0.93 & 0.41\\
\hline
$M_{sat}$ ($10^{10} M_\odot$) & 2.20 & 0.21 & 0.21 & 0.14\\
\hline
$\epsilon$ (kpc) & 0.1 & 0.2 & 0.2 & 0.1\\
\hline
$N_{host}$ &100000  &48000  & 25714  &24000  \\
\hline
$N_{sat}$ &33333 & 32000 & 5714  &8000 \\
\hline
\end{tabular}
\label{table1}
\end{center}
\end{table}
%\caption{Mass particles (Mass), gravitational softening lengths ($\epsilon$) and
%number of particles ($N$) used for each component of the galaxies.}
%\end{table}

The cooling and star formation (SF) recipes follow the
prescriptions of Thomas \& Couchman (1992) and of Katz et al. (1996),
respectively.
 Gas particles with $T>10^4$K
cool at constant density
%(with the assumption of solar metallicity)
 for the duration of a timestep. Gas particles with
$T< 2\times 10^4 K$, number density $n > 0.1\, cm^{-3}$,
overdensity $\Delta \rho_{gas}> 100$ and ${\bf \nabla . \upsilon}
<0$ form stars according to the standard SFR prescription:
$d\rho_*/dt = c_* \rho_{gas}/t_{dyn}$, where $\rho_*$ refers to
the stellar density, $t_{dyn}$ is the dynamical
timescale of the gas and $c_*$ is the SF efficiency.
%Instead of creating new (lighter) star particles, we implement the SF
%prescription in a probabilistic fashion.
Assuming a constant
dynamical time across the timestep, the fractional change in
stellar density is $\Delta \rho_*/\rho_* = 1-\exp(-c_* \Delta
t/t_{\rm dyn})$. For each gas particle, we draw a random number
($r$) between 0 and 1 and convert it to a star if $r<\Delta
\rho_*/\rho_*$.

%The energy injection into the inter-stellar medium (ISM) from SN,
%which regulates the star formation rate (SFR), is modelled following
%the approach of Durier \& de Freitas Pacheco (2008, in prep.). 
Instead of assuming `instantaneous' energy injection, we include the effective
lifetime  of SN progenitors using the rate of energy injection
$H_{SN}$. For this, we consider stellar lifetimes in the mass
ranges $0.8\,M_\odot\,<m<8.0\,M_\odot$ and
$8.0\,M_\odot<m<80.0\,M_\odot$ for Type Ia and Type II progenitors
respectively. Using a Salpeter initial mass function for Type II
SN gives:

\begin{equation}
H_{SN_{II}}=2.5\times10^{-18}\Big(\frac{m_*}{M_\odot}\Big)E_{SN}\Big(\frac{1300}{\tau(\textnormal{Myr})-3}\Big)^{0.24}
\textnormal{erg.s$^{-1}$},
\end{equation}

\noindent
where $E_{SN}=10^{51}$ erg, $m_*$ is the mass of the stellar
population and $3.53 <\tau <29$ Myr. For Type Ia SN, the heating
is delayed, since they appear $t_0=0.8-1.0$ Gyr after the onset of
star formation. Following de Freitas Pacheco (1998), the
probability of one event in a timescale $\tau$ after the onset of
star formation is given by:

\begin{equation}
H_{SN_{I_a}}=4.8\times10^{-20}\Big(\frac{m_*}{M_\odot}\Big)E_{SN}\Big(\frac{t_0}{\tau}\Big)^{3/2}
\textnormal{erg.s$^{-1}$}.
\end{equation}

Eqns (1) and (2) are used to compute the energy released
by SN derived from a star particle $i$ ($E_i$). A fraction $\gamma$ of
this energy is deposited in the j$^{th}$ neighbour gas particle by
applying a radial kick to its velocity with a magnitude $\Delta
v_j = \sqrt{(2w_j\gamma E_i/m_j)}$, where $w_j$ is the weighting
based on the smoothing kernel and $m_j$ is the mass of gas
particle j. We note that all gas neighbours are located in a
sphere of radius $R_{SN}$, centered on the SN progenitor, to avoid
spurious injection of energy outside the SN's region of influence.
In the following, we use the following standard values: $\gamma=0.1$,
$R_{SN}=0.4$ kpc and $c_* = 0.01$.
% which are quite standard values 
%(see for instance Stinson et al. 2006).
% When galaxies are isolated, these parameters lead to a quasi-constant
%SFR  of $2 M_\odot yr^{-1}$ and
% are in good agreement with those for low-mass objects
%found in previous simulations of isolated galaxies
%(see for instance Stinson et al. 2006).
 
\subsection{Results}

\begin{figure}
\rotatebox{0}{\includegraphics[width=4.4cm]{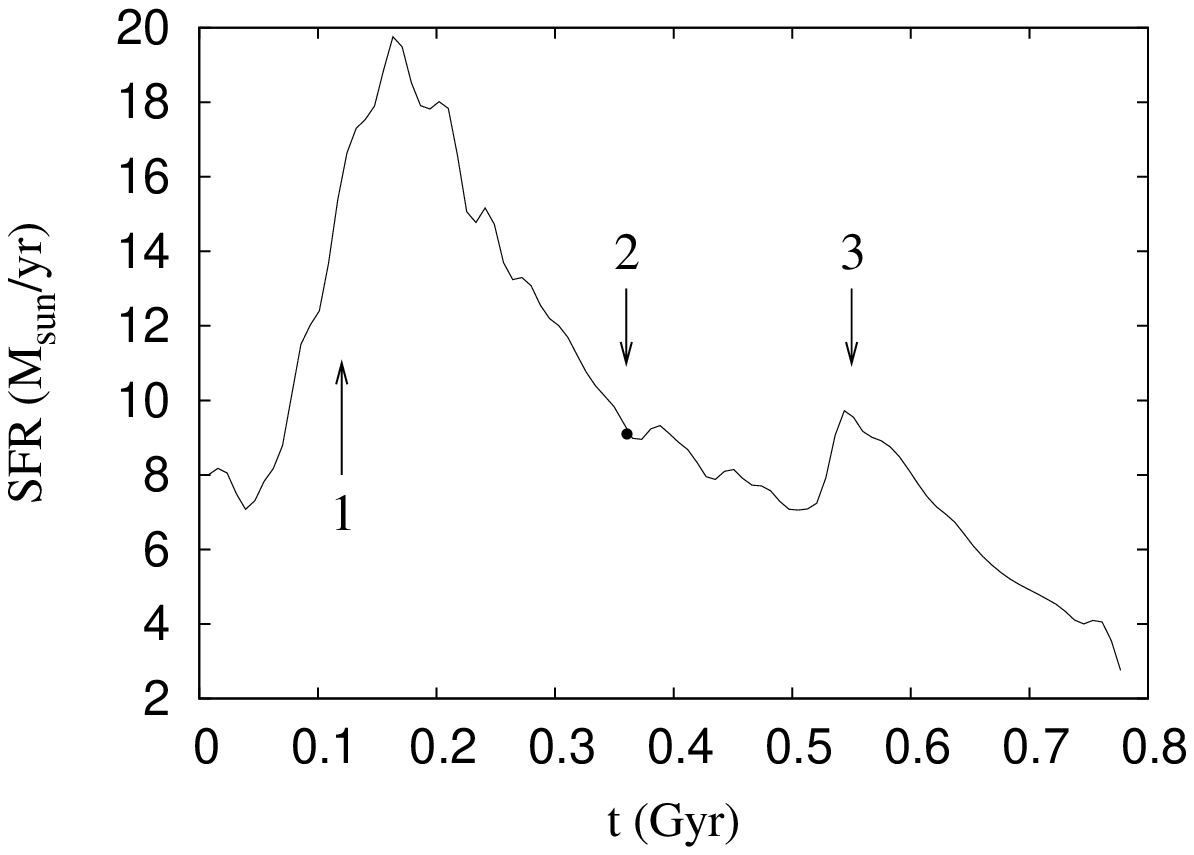}}
\rotatebox{0}{\includegraphics[width=4.4cm]{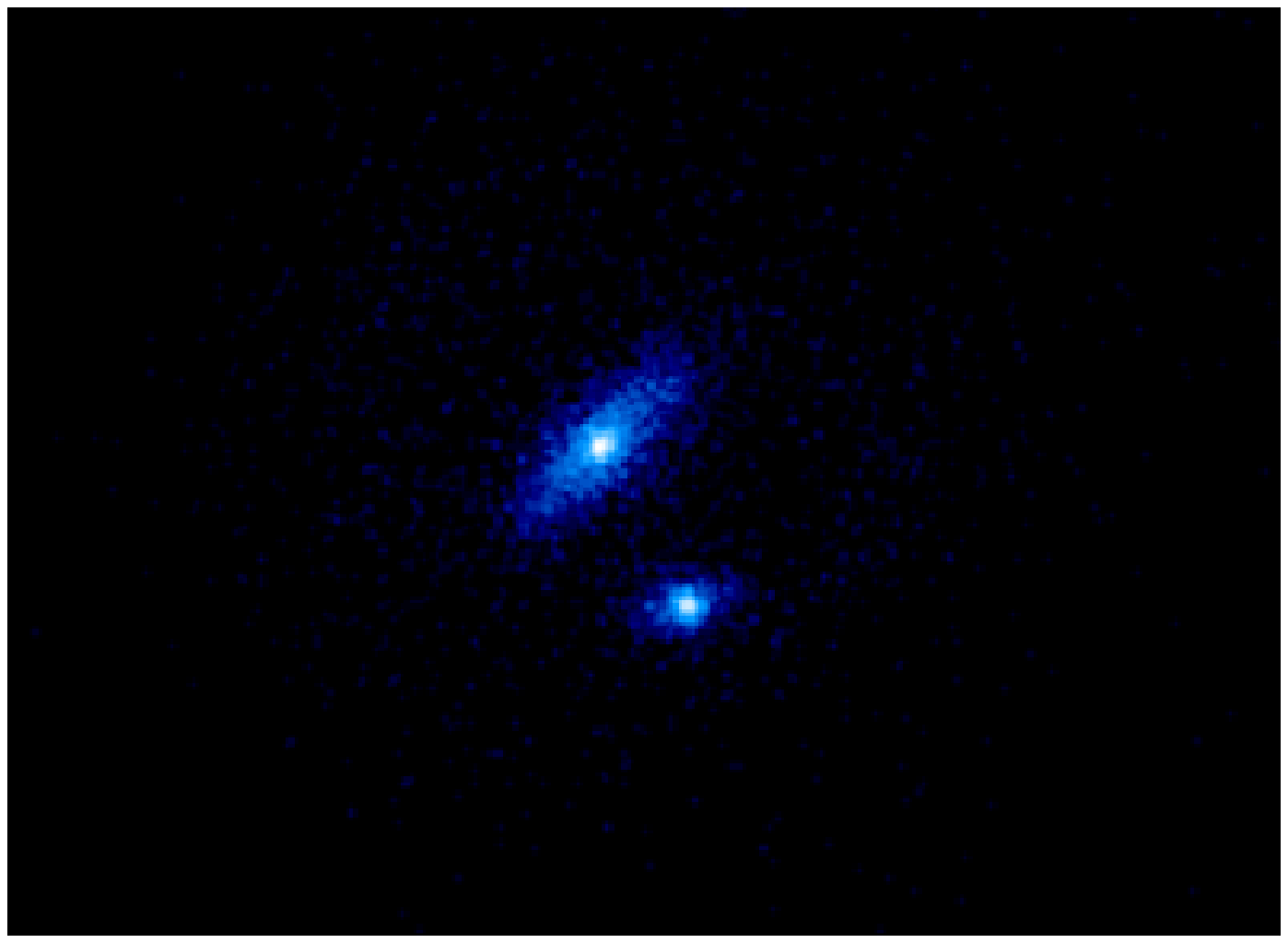}}
\caption{The star formation rate obtained from our model (left panel)
 and projected distribution of newly formed stars at $t=0.36$ Gyr
 (right panel). The arrows in the left panel correspond to the SFR at
 the first encounter (1), at $t=0.36$ Gyr (2) and at the final plunge (3).
} 
\label{sfr}
 \end{figure}

Figure (\ref{simulated_galaxy}) shows the time evolution of the
projected stellar number density of the system. After $0.36$ Gyr 
 from the beginning of the simulation, 
the system has a stellar mass of $\sim 1.92 \times 10^{10} M_{\odot}$ 
and presents, from a morphological point of view, 
a general shape and patterns similar to the observed
ones. The disc of the more massive galaxy has not been destroyed
 yet, reflecting the very early stage of the dynamical process,
 and
is viewed in Figs~\ref{simulated_galaxy} and \ref{sfr} at an
inclination of $\sim 26^\circ$, as 
found from its outer isophote. The bar of the host galaxy has roughly
the right orientation and length ($\sim 5$ kpc) and the projected position of
the satellite remnant is in rough agreement. However, many
difficulties remain e.g. in reproducing the blue arm or tidal tail
on the upper right and the fact that the companion should be split
into two distinct bright knots.

In the present scenario, the bar is forming after the
first encounter with the satellite, since at that time there is a very
strong triggering. In good agreement with the observations, it is
particularly clear in the 
newly formed stars, since those, being on near-circular orbits, are
more prone to the bar instability. Also in good agreement is the fact that
the newly formed stars are also located in the satellite remnant
(fig. \ref{sfr}). Moreover, according to our model,
J033239.72-275154.7 is observed when the satellite is about to have the second
encounter with the host. 
At this specific time, the SFR is $\sim 9 M_\odot.yr^{-1}$ (fig. \ref{sfr})
in good agreement with the observational estimation.

%%%%%%%%%%%%%%%%%%%%%%%%%%%%%%%%%%%%%%%%%%%%%%%%%

\begin{figure}
\rotatebox{0}{\includegraphics[width=4.4cm]{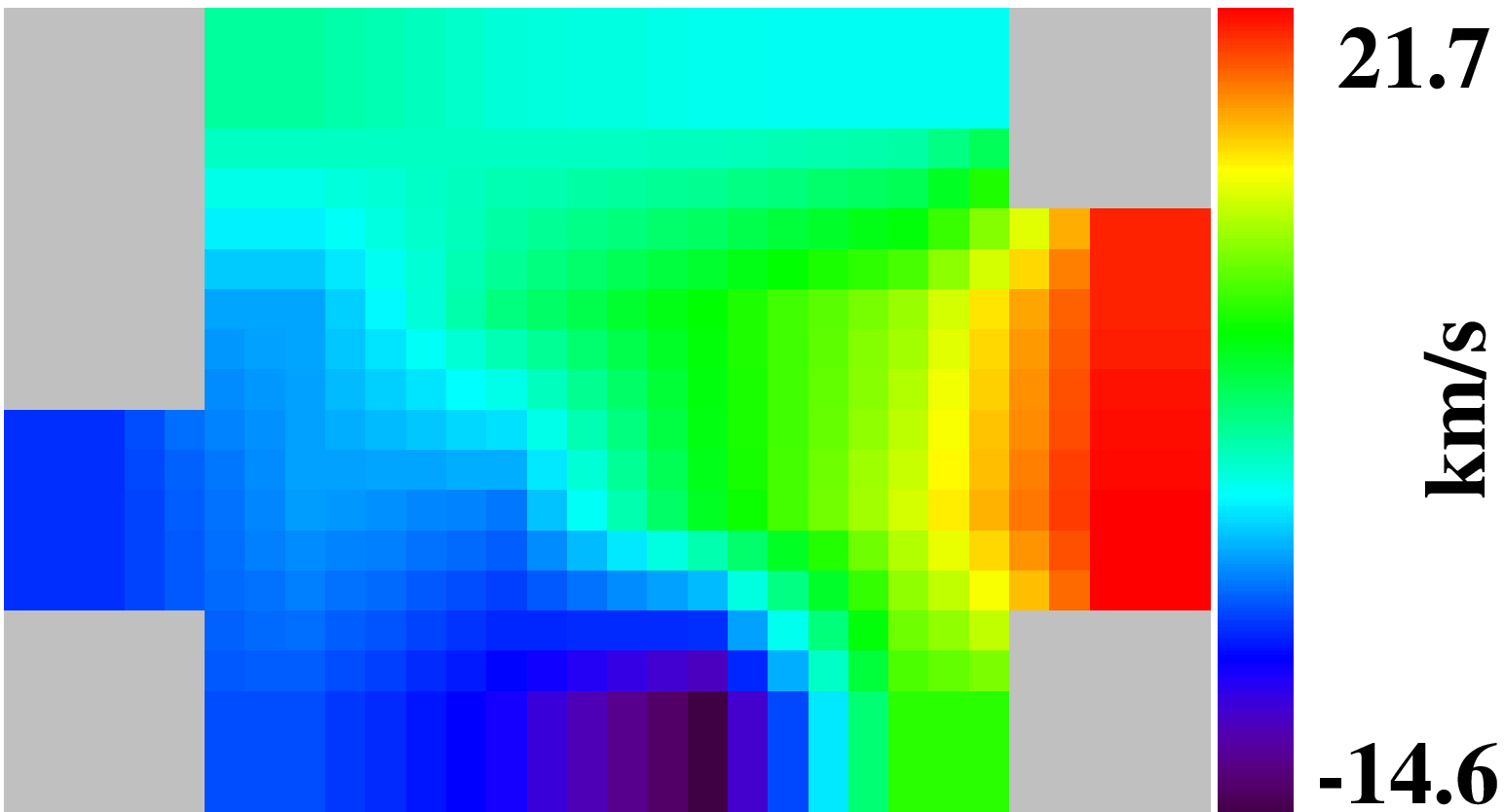}}
\rotatebox{0}{\includegraphics[width=4.4cm]{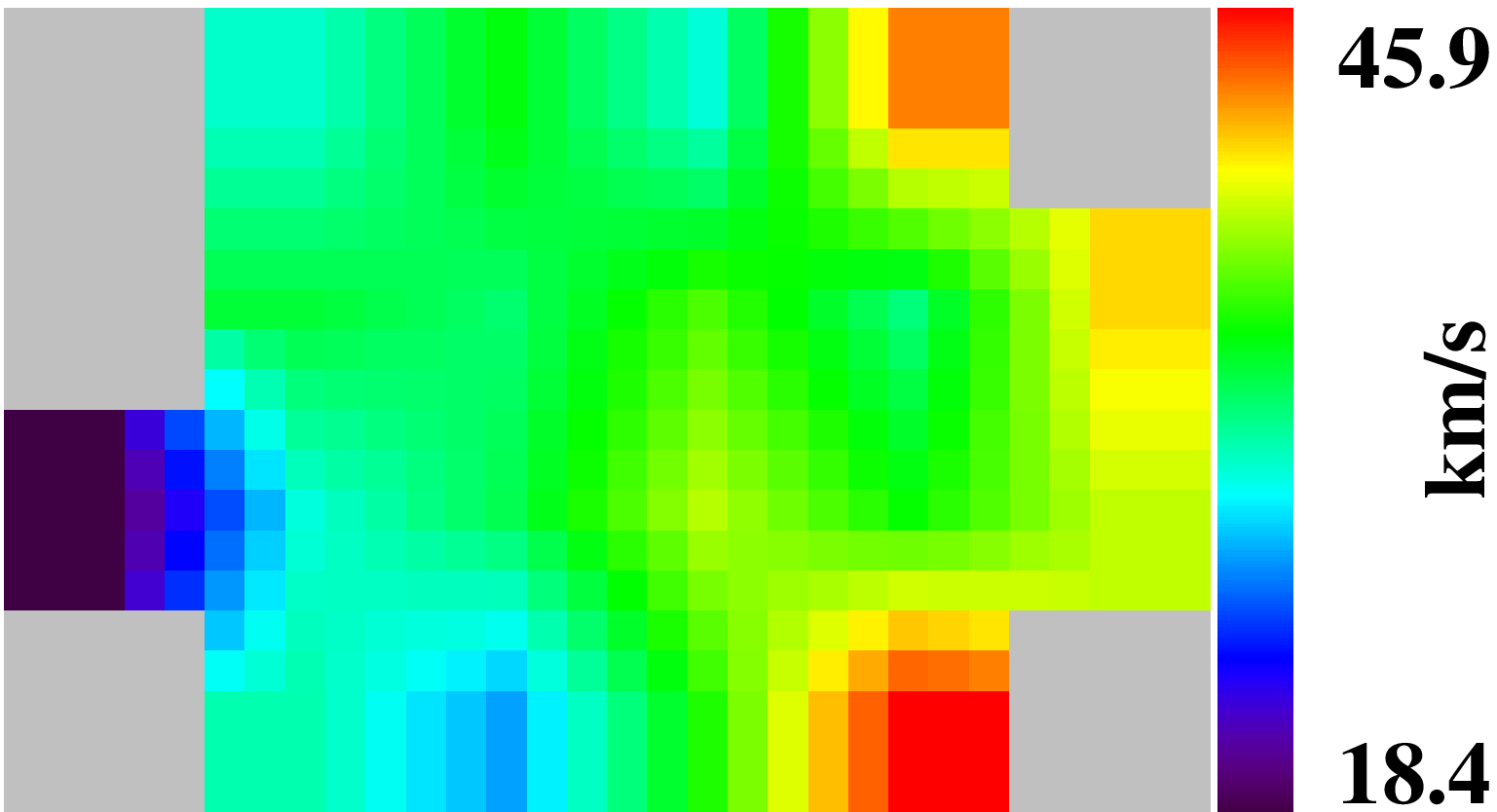}}
\caption{Maps of the velocity field (on the left) and $\sigma$ field 
 (on the right) of the gas component obtained from our simulation.}
 \label{v_s_sim}
 \end{figure}

The VF and $\sigma$ map of the gas component derived from our numerical model are   
represented in the  GIRAFFE format in Fig.\ref{v_s_sim}.
The determination of the radial velocities has been realized
 at the orientation  of Figs \ref{simulated_galaxy} and \ref{sfr}.
The observed VF and its  amplitude are faithfully reproduced and show that  
the initial inclination 
%of $15^\circ$
 between the two orbital planes is consistent.
% has proved to be a good choice to match the maximal
%amplitudes of the velocity field.
The $\sigma$ map is relatively flat around a mean value of $35-40$
km/s, in good agreement with the observations. It presents two maxima
of roughly the correct value, but located at positions other than
those observed. However, these two maxima are located in regions 
 which are not well resolved in gas particles and thus the resulting
pixel values are not significant.

\begin{figure}
%\rotatebox{0}{\includegraphics[width=\columnwidth]{max.eps}}
\rotatebox{0}{\includegraphics[width=\columnwidth, height=6cm]{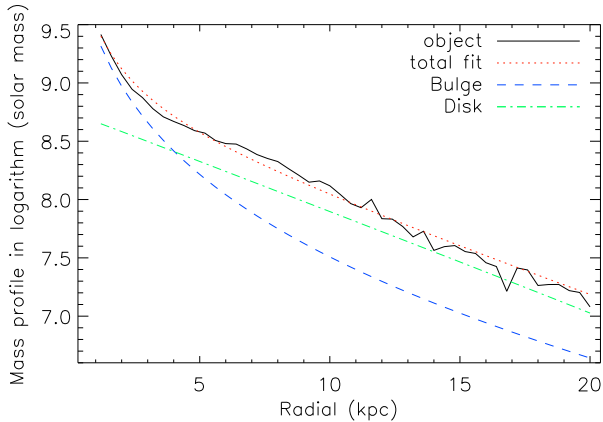}}
\caption{
Mass profile decomposition along the major axis
of the final product. The mass profile of the final
product is show by the black line. The green dotted-dashed line,
the blue dashed line and the red dotted line indicate
the disk component, the bulge component and the sum of them,
respectively.
}
 \label{bulk2disk}
 \end{figure}

After 4.26 Gyrs, corresponding to the present time $z=0$, the final object is
characterized by a bulge-to-disk ratio close to 1.04,
as derived from the mass profile decomposition along the major axis
of the final product (see Fig. \ref{bulk2disk}).
For this decomposition, we used standard models for the disk and bulge,
 namely an exponential disk (S\'ersic profile with index $n=1$) and a bulge following
 de Vaucouleurs law (S\'ersic profile with index $n=4$),
 in which
the free parameters are the bulge and disk radius and
the bulge and disk flux. 
The rotational support parameter
$V_{maj}/\sigma_c=1.62$, where $V_{maj}$ is the maximum of the major axis
rotation speed and $\sigma_c$ the central velocity dispersion.  This
latter is estimated by considering all star particles in a sphere of
radius 0.5 kpc.
These previous values obtained highly suggest the formation of a S0 galaxy.

%%%%%%%%%%%%%%%%%%%%%%%%%%%%%%%%%%%%%%%%%%%%%%%%%
%                   FIGURE TWO
%%%%%%%%%%%%%%%%%%%%%%%%%%%%%%%%%%%%%%%%%%%%%%%%%
%%%%%%%%%%%%%%%%%%%%%%%%%%%%%%%%%%%%%%%%%%%%%%%%%%

%%%%%%%%%%%%%%%%%%%%%%%%%%%%%%%%%%%%%%%%%%%%%%%%%
%                   FIGURE TWO
%%%%%%%%%%%%%%%%%%%%%%%%%%%%%%%%%%%%%%%%%%%%%%%%%
%%%%%%%%%%%%%%%%%%%%%%%%%%%%%%%%%%%%%%%%%%%%%%%%%%

%\vspace{-0.2cm}
\section{Discussion and Conclusions}

In this paper, we presented photometric and kinematic data of
J033239.72-275154.7 and made a detailed comparison with an $N$-body
simulation. 
%Observational aspects have been firstly studied by Yang
%et al. (2008) in which the morphological and
%kinematic properties of this object have been derived using the GIRAFFE spectrograph.
%As far as the numerical modelling is concerned, we ran a large number of N-body
%simulations of the merger of two spiral galaxies including  star formation prescription.
A first interesting and important result is that among all the simulations
that we have realized  with specific
orbital parameters and progenitor mass ratios, the combination of
parameters such as 
the star formation history, the VF and $\sigma$ map put strong constraints on models
and can remove the degeneracy of good candidates based on morphological criteria only.

We found that our fiducial model can roughly reproduce the general
morphological shape and the total stellar mass of 
the object as well as the observed SF rate, VF and intensity
of the $\sigma$ map.
This specifically includes the reproduction of the giant young bar,
its location and shape, the relative projected location of the
companion, the
overall morphological shape of the galaxy, its low rotation
and its off-center dynamical axis. The model predicts a slow-down
of the velocity of the host galaxy due to exchanges of angular 
momentum with the companion  (see Fig. \ref{simulated_galaxy}). 
However, using such idealized simulations, we were unable to reproduce
many features, such as the blue arm or tidal tail on the upper right part, the
morphological patterns of the two bright knots of the satellite remnant 
and the location of the $\sigma$ peaks. Such discrepancies with
observations can be easily understood as due to the huge 
parameter space to be covered by the simulations. These include the
properties of the progenitors (halo:disk:bulge:gas mass ratios and
relative extents, shape of their density profiles and kinematics),
the geometry of the encounter and of the viewing, the gas properties,
the SF, cooling and feedback modeling.
On the other hand, the discrepancies with observations may also have
astrophysical origin and/or be due to a more complex formation
scenario. Several such possibilities -- such as multiple encounters or
progenitors with specific properties -- come to mind, but would
increase substantially the already too large free parameter space.

  In  spite of these difficulties, it's encouraging to note that our
simple numerical modelling is able to build a consistent picture of
the formation of J033239.72-275154.7. 
This system is consistent with having been formed
from a merger of two objects with a mass ratio 1:3. The simulation indicates
that a bar is forming in the host galaxy after the first passage of the satellite
where an important fraction of available gas is consumed in the induced
burst giving a plausible explanation of  the observed blue colors of
the bar and of the satellite remnant. 
Moreover, both VF and $\sigma$ map derived from the simulation match with the
observational values and thus support this scenario.
In its later evolution, we found that J033239.72-275154.7
may become a S0 galaxy, as suggested by the results of our simulations.
This is mainly explained  by the fact that the host galaxy 
experiences a violent relaxation and
looses some
angular momentum during the merger process due to the retrograde orbit
of the satellite. Moreover,
by loosing some angular momentum, some of the gas can sink toward the
center of the galaxy where it can be converted into new stars and then
accelerate the growth of the bulge.

To finish, it is interesting to point out that
 recent numerical studies have demonstrated that gas-rich mergers can produce remnant
disks provided strong feedback processes and also if both stellar and gas
component do not experience a significant angular momentum loss 
(Springel \& Hernquist 2005; Robertson et al. 2006; Governato et
 al. 2007, Hopkins et al. 2008).
We propose in this work an extreme example where i) the initial orbital
configuration of the merger event do not permit to satisfy the latter
 criteria  and ii) the progenitors of the system (in our
 numerical model) may 
not be sufficiently gas rich to reform 
a significant disk according those past studies. 
However, the present results are  consistent with  the previous ones 
in the emphasis on the fundamental role
played by the  last major event in building the Hubble sequence
(see e.g. Hammer et al., 2005 and 2007).
% since from a $S_a$ or
%$S_b$ we obtained a  $S_0$.
It would thus be interesting to determine the frequency
of each orbital configuration during a merger event and to compare with
the galaxy population in the
local universe to confirm this galaxy formation scenario.

\begin{acknowledgements}
S. P.  acknowledges the financial support through a ANR grant.
E. A. from ANR-06-BLAN-0172. 
We warmly thank the referee Brant Robertson for an insightful
review that considerably improved the quality of the original
 manuscript.
S. P. also warmly  thank Y. Kakazu for
interesting conversations and for her moral support.

\end{acknowledgements}

\end{document}